\documentclass[aps,pra]{revtex4-1}
\usepackage{graphicx, amsmath, amsfonts, amssymb}
\usepackage{array}
\usepackage{rotating}
\usepackage{graphicx}

\begin{document}
\title{Pair states in one-dimensional Dirac systems}

\author{R. R. Hartmann}
\email[]{richard.hartmann@dlsu.edu.ph}
\affiliation{
Physics Department,
De La Salle University,
2401 Taft Avenue,
0922 Manila,
Philippines.
}

\author{M. E. Portnoi}
\email[]{m.e.portnoi@exeter.ac.uk}
\affiliation{School of Physics, University of Exeter, Stocker Road, Exeter EX4 4QL, United Kingdom\\
and International Institute of Physics, Universidade Federal do Rio Grande do Norte, Natal - RN, Brazil.
}

\begin{abstract}
Analytic solutions of the quantum relativistic two-body problem are obtained for an interaction potential modeled as a one-dimensional smooth square well. Both stationary and moving pairs are considered and the limit of the delta function interaction is studied in depth. Our result can be utilized for understanding excitonic states in narrow-gap carbon nanotubes. We also show the existence of bound states within the gap for a pair of particles of the same charge.
\end{abstract}
\maketitle

\section{INTRODUCTION}
The analytical solutions supported by the square well and Delta potential for the Schr{\"o}dinger equation make them among the most widely used potentials in non-relativistic quantum theory and they have given fundamental insights in the field of theoretical physics. In the relativistic case, potentials containing discontinuities or kinks are often harder to treat, since the relationship between the two components of the wavefunction leads to the derivative of the potential entering the wave equation. Nevertheless, the square well has been used with great success to describe single-particle phenomenon in quasi-relativistic one and two-dimensional systems \cite{katsnelson2006chiral,neto2009electronic,PhysRevB.74.045424}. Here, we provide a way to obtain the square well solutions for the quasi-one-dimensional two-body relativistic problem, from the smooth analytically solvable barrier side, which foregoes the need to consider special boundary conditions. With the recent surge of Dirac materials \cite{wehling2014dirac} has come a renewed interest in many quasi-relativistic phenomena and table top experiments now offer the possibility to check many relativistic theories.  Currently the experimental focus is on attractive potentials to reveal the role of excitonic effects, but we show that within the same formalism, quasi-one-dimensional systems can also support bound states within the band gap for two repelling particles. In light of the ongoing extensive search of new-quasi particles in one and two-dimensional systems, such as Majorana fermions, our result for binding same-charged particles is of significant potential importance. Indeed, of special interest is the case where both the electron-electron and the electron-hole pair binding energy correspond to the middle of the band gap, as it enforces electron-hole symmetry without the recourse to a superconductor. This state is also energetically favorable as it reduces the Fermi energy of a doped system.

Relevant examples of quasi-one-dimensional relativistic systems, for which our results are applicable, are narrow-gap carbon-nanotubes and graphene nanoribbons. For these carbon-based nanostructures, the interaction potential between a pair of particles may be considered as a quasi-one-dimensional problem, averaged over the nanotube diameter or nanoribbon width. For these systems, the interaction potential is flat-bottomed as it varies very little over the averaging scale. In many carbon nanotubes devices, a metallic substrate is used as the gate electrode to manipulate the fermi-level \cite{nygaard2000kondo}. The presence of the substrate results in image charges, which make the interaction decay faster than the widely employed Coulomb potential \cite{PhysRevA.90.052116}, therefore the flat-bottom potential is a reasonable interaction model for a carbon nanotube or graphene nanoribbon above a metallic gate.

Charge carriers in graphene, a single monolayer of carbon in a honeycomb lattice \cite{neto2009electronic}, are described by the same equation used to describe
two dimensional massless Dirac-fermions, the Dirac-Weyl's equation. By using a simple tight-binding model, Wallace \cite{wallace1947band} demonstrated that a pristine graphene sheet has no band gap, and that the conduction and valence bands are linear near the crossing points. A single-walled nanotube can be thought of as a graphene sheet rolled into a seamless cylinder. In the absence of curvature effects, within the frame of a simple tight-binding model, this rolling can result in either a semi-conducting or metallic tube \cite{SaitoDresselhaus98Book}. However in reality all but armchair nanotubes are metallic,  since non-zero curvature \cite{PRLKane97,ANDO_JPSJ.69.1757,PhysRevB.63.073408,ANDO_JPSJ.74.777,hartmann2015terahertz} gives rise to small bandgaps which can be of the order of a few meV, corresponding to terahertz (THz) frequencies. The size of these gaps can be tuned by application of a magnetic field along the nanotube axis \cite{AjikiAndo93JPSJ,ReichBook04,KonoRoche06CRC,KonoetAl07Book}. The diverse applications of THz radiation and its importance to fundamental science makes finding ways to generate, manipulate, and detect it one of the foremost challenges in modern applied physics \cite{LeeWankeAl07S}. One approach to fill the THz gap is to utilize narrow-gapped single-wall carbon nanotubes \cite{hartmann2014terahertz}. These tubes can exhibit strong THz optical transitions, which can be manipulated via externally applied magnetic and electric fields, giving rise to the possibility of utilizing them as highly tunable, optically-active materials in THz devices \cite{kibis2007generation,portnoi2008terahertz,portnoi2009magnetically,batrakov2010terahertz,Batrakov_Physica_B_2010,hartmann2014terahertz,hartmann2015terahertz}.

In semiconductor carbon nanotubes, the existence of low-lying dark excitons drastically suppresses the photoluminescence efficiency \cite{ando1997excitons,WangetAl05Science,MaultzschetAl05PRB2,Kono_07_LPR}. However, in narrow-gap nanotubes it has been demonstrated numerically, that the binding energy of certain potentials scale with the bandgap. Therefore, undesirable effects due to dark excitons should not dominate optical processes in narrow-gap nanotubes \cite{hartmann2011excitons,portnoi2013excitons}. However, the question of exactly how excitonic effects influence the optical processes in narrow-gap nanotube is still an outstanding problem. Excitonic effects in Dirac materials have been studied using a variety of approaches such as the Bethe-Salpeter method \cite{wang2011excitonic} as well as the two-body matrix Hamiltonian, based on the low energy expansion of the tight-binding \cite{sabio2010two,hartmann2011excitons,lee2012quasilocalized,downing2015bielectron,sablikov2017two}, which shall be the method employed in this study. Pair formation has been studied in Dirac materials with effective mass, such as gapped graphene \cite{iyengar2007excitations,berman2008collective,sabio2010two,berman2013coupling,PhysRevBEgger97}, bilayer graphene \cite{zarenia2013electron}, as well as graphene in the trigonal warping regime \cite{mahmoodian2012moving,mahmoodian2013moving,marnham2014metastable}. However, there is still much debate concerning the existence of coupled pairs in intrinsic graphene \cite{yang2011excitons,ratnikov2012size,downing2015bielectron}. The two-body problem has also been the subject of study in narrow gap carbon nanotubes and graphene nanoribbons, and some analytic solutions have been found \cite{hartmann2011excitons,portnoi2013excitons,HartmannAIP,Monozon201689}. Numerical methods have also been used to determine exciton bound states in metallic carbon nanotubes subjected to a magnetic field \cite{ando1997excitons}. Previous analytic results for finite potentials were limited to mid-gap states, relying on numerical methods to determine the remaining spectrum.  Unlike previous studies we calculate the full positive energy spectrum exactly for a flat-bottomed potential, offering a powerful tool for modeling exciton energy levels.

The eigenvalues of a non-relativistic particle, subjected to a confining potential, are obtained by first solving the Schr{\"o}dinger equation and then imposing the appropriate boundary conditions upon the wave function. For the Schr{\"o}dinger equation, the simplicity of the infinite square well offers many insights into quantum effects and serves as a useful approximation for more complex quantum systems. Contrastingly, the solution of the particle in a box problem in the relativistic regime is certainly non-trivial \cite{klein1929reflexion,afanasiev1990does,macia1991scattering,alhaidari2009dirac, alonso1997boundary,alonso1997general,alberto1996relativistic}. After solving the Dirac equation and requiring that all the components of the spinor vanish at the wells edge, the only permissible solution is the null wave function, the same is true for the cylindrical infinite well \cite{percoco1989aharonov,chen2007fock,matulis2008quasibound}. Supplementary boundary conditions may be employed to resolve this problem such as introducing a mass going to infinity outside of the well \cite{alberto1996relativistic,thaller1992dirac,volkov2009electrons}, such boundary conditions can relax the continuity of the wave function at the well's boundary yet preserve the continuity of the probability density across the well. However, different forms of quantum impenetrability may lead to different physical consequences \cite{afanasiev1990does}. Similar problems arise for the relativistic two-body problem. A discontinuous potential imposes many restrictions on the spinor components which are often very difficult to satisfy, and demanding that the wave function and its derivative are continuous at the boundary results in the null wave function. To avoid the boundary condition issues we consider a smooth piece-wise step potential, that contains an adjustable parameter, which can be varied such that in some limit the potential transforms into the Heaviside step function. Therefore, by symmetry one can construct a truly flat bottomed potential, which can be solved without the need of invoking supplementary boundary conditions. Exact solutions of the Dirac equation are not only useful in the analytic modeling of physical systems, but they are also important for testing numerical, perturbation or semi-classical methods. The smooth square well gives valuable insights in the behavior of two Dirac-like particles interacting via a short-range potential.

In what follows we consider excitons formed by relativistic one-dimensional electrons and holes interacting via a flat bottomed piece-wise potential. We first focus on excitons possessing zero total momentum along the nanotube axis. A solution to the two-body Dirac problem in the rest frame, for a smooth step potential is presented. The wavefunctions are expressed in terms of Heun confluent functions and all the spinor components and their derivatives are continuous throughout space. The solutions are then analyzed in the limit in which the potential transforms into a true step potential and via symmetry conditions the quantized energy spectrum of the square well is attained. This potential is then used to model the interaction potential between an electron and hole in a one-dimensional Dirac system and the binding energy is shown to scale with the band gap. The true square well problem is then revisited. By analyzing the wave function of the smooth square well, appropriate boundary conditions are obtained for the true square well. This enables one to obtain the energy spectrum of an exciton possessing finite total momentum along the nanotube axis. Lastly, the model potential is analyzed in the delta function limit, for both the non-relativistic and relativistic regime.




\section{Solution of the flat-bottom interaction potential problem for two Dirac particles}
In the absence of curvature effects, the single-particle Hamiltonian of a nanotube may be obtained from graphene by applying the periodic boundary condition along the direction of the circumference. Curvature effects are equivalent to introducing an effective flux along the tube, which is equivalent to shifting the momentum of charge carriers in an unrolled graphene sheet. Therefore, the effect of curvature and applied magnetic fields are directly analogous to the one-dimensional graphene problem where the curvature induced gap plays the role of fixed transverse momentum.
In general, the single-particle Hamiltonian of a narrow gap carbon nanotube, of bandgap $2\hbar v_{{\rm F}} \kappa _{y}$, is given in the vicinity of the bandgap edge by $\hbar v_{{\rm F}} \left(\sigma _{x} \hat{\kappa }+\sigma _{y} \kappa _{y} \right)$, where $\sigma _{x,y}$ are the Pauli spin matrices, $v_{{\rm F}} $ is the Fermi velocity in graphene, and $\hat{\kappa }$ is the operator of the wave vector along the nanotube axis (x-axis). The corresponding eigenvalues are given by $\varepsilon =\pm \hbar v_{{\rm F}} \sqrt{\kappa ^{2} +\kappa _{y}^{2} }.$ For an electron-hole pair, the Hamiltonian can be written as \cite{hartmann2011excitons}:
\begin{equation}
\hat{H}=\hbar v_{{\rm F}} \left(\begin{array}{cccc} {0} & {\hat{\kappa }_{e}^{} -i\kappa _{y} } & {-\hat{\kappa }_{h}^{} +i\kappa _{y} } & {0} \\ {\hat{\kappa }_{e}^{} +i\kappa _{y} } & {0} & {0} & {-\hat{\kappa }_{h}^{} +i\kappa _{y} } \\ {-\hat{\kappa }_{h}^{} -i\kappa _{y} } & {0} & {0} & {\hat{\kappa }_{e}^{} -i\kappa _{y} } \\ {0} & {-\hat{\kappa }_{h}^{} -i\kappa _{y} } & {\hat{\kappa }_{e}^{} +i\kappa _{y} } & {0} \end{array}\right),
\end{equation}
where the indices $e$ and $h$ correspond to the electrons and holes with $\hat{\kappa }_{e,h} =-i\partial /\partial x_{e,h}$, where $x_{e}$ and $x_{h}$ are the positions of the electron and hole along the nanotube. This Hamiltonian acts on the basis $|\Psi _{ij} \rangle =|\psi _{i}^{e} \rangle |\psi _{j}^{h} \rangle $, where the indices $i$ and $j$ correspond to the carbon atoms of the two different sub-lattices in the honeycomb lattice. In the absence of interaction and band-filling effects this Hamiltonian yields four energy eigenvalues corresponding to a pair of non-interacting quasi-particles:
\begin{equation}
\varepsilon =\hbar v_{{\rm F}} (\pm \sqrt{\kappa _{y}^{2} +\kappa _{e}^{2} } \pm \sqrt{\kappa _{y}^{2} +\kappa _{h}^{2} } ).
\label{4energy}
\end{equation}
In this formalism, when considering a system containing a single electron and a single hole, one should only consider the solution with positive signs and the band gap of the two-particle system is given by $E_{g} =2\hbar v_{{\rm F}} \kappa _{y}$. It should also be noted, that there also exists a possibility of binding a pair of same charged particles. This case corresponds to the solutions with negative signs, interacting via a repulsive potential. 
In what follows, we shall restrict ourselves to the single valley regime. However, the complete treatment of the problem requires that all valley and spin quantum numbers be taken into account, in this instance the number of different types of excitons associated with a given carbon nanotube spectrum branch rises to 16 \cite{ando2009theory}. The full treatment of the problem shall be the topic of future study.

The interaction potential $U\left(x_{e} -x_{h} \right)$ is a function of the relative separation between the particles only, therefore it is convenient to move to the center of mass and relative motion coordinates: $X=\left(x_{e} +x_{h} \right)/2$, $x=x_{e} -x_{h} $. Therefore, the operators can be expressed as $\hat{\kappa }_{e}^{} =\hat{K}/2+\hat{k}$ and $\hat{\kappa }_{h}^{} =\hat{K}/2-\hat{k}$ where $\hat{k}=-i\partial /\partial x$. The wave function of the interacting particles can by written as $\Psi _{ij} \left(X,{\kern 1pt} x\right)=e^{iKX} \phi _{ij} \left(x\right)$ allowing the operator $\hat{K}$ to be replaced with the constant $K$ which represents the wave vector of the interacting particles center of mass. Upon separating relative and center of mass motion it is more convenient to move to the symmetrized wave functions:
\begin{equation}
\begin{array}{cc}
\psi _{1} =\phi _{{\rm BA}} -\phi _{{\rm AB}}, & \psi _{2} =\phi _{{\rm AA}} -\phi _{{\rm BB}},\\
\psi _{3} =\phi _{{\rm AA}} +\phi _{{\rm BB}}, & \psi _{4} =\phi _{{\rm BA}} +\phi _{{\rm AB}},
\end{array}
\end{equation}
This enables the eigenvalue problem to be expressed as
\begin{equation}
\hbar v_{F}\hat{M}_{nm}\psi_{m}=\left[\varepsilon-U\left(x_{e}-x_{h}\right)\right]\psi_{n}
\label{GrindEQ1}
\end{equation}
where
\begin{equation}
\hat{M}=\left(
\begin{array}{cccc} {0} & {K} & {i2\kappa _{y} } & {0} \\ {K} & {0} & {0} & {0} \\ {-i2\kappa _{y} } & {0} & {0} & {2\hat{k}} \\ {0} & {0} & {2\hat{k}} & {0}
\end{array}\right).
\end{equation}
Let us first consider the case of the stationary exciton, which couples to light, i.e. $K=0$. In this instance, $\phi _{{\rm AA}} =\phi _{{\rm BB}} $,  which allows us to reduce Eq. (\ref{GrindEQ1}) from a system of four equations down to three. Equation (\ref{GrindEQ1}) can be reduced to a single second order equation in $\psi _{3}$:
\begin{equation}
\frac{\partial^{2}\psi_{3}}{\partial z^{2}}-\frac{1}{\left(E-V\right)}\frac{\partial\left(E-V\right)}{\partial z}\frac{\partial\psi_{3}}{\partial z}+\frac{1}{4}\left[\left(E-V\right)^{2}-4\Delta^{2}\right]\psi_{3}=0,
\label{GrindEQ2}
\end{equation}
where we have scaled the eigenvalue $E=\varepsilon L/\hbar v_{{\rm F}}$,  potential energy $V=UL/\hbar v_{{\rm F}} $, momentum $\Delta =\kappa _{y} L$ and made use of the variable change $z=\left(x-W/2\right)/L$ where $W$ is the effective width of the well and $L$ a constant. The remaining components $\psi _{1} $ and $\psi _{4} $ are found via the relations:
\begin{equation}
\psi _{1} =i\frac{2\Delta }{\left(E-V\right)} \psi _{3},
\label{eq:pairing1}
\end{equation}
\begin{equation}
\psi _{4} =-i\frac{2}{\left(E-V\right)} \frac{\partial \psi _{3}}{\partial z},
\label{eq:pairing2}
\end{equation}
Since our primary interest is the study of optoelectronic applications of carbon nanotubes \cite{hartmann2014terahertz} we shall restrict ourselves to calculations concerning electron-hole pairs. However, the approach used here can be easily generalized for the study of same-charge particle pairs \cite{downing2015bielectron,sablikov2017two}. Indeed, it can be seen that exchanging $E$ to $-E$ and $V(x)$ to $-V(x)$ leaves equation Eq.~(\ref{GrindEQ2}) unchanged. Therefore, proving the existence of electron-hole pairs interacting via an attractive potential within the gap also demonstrates the existence of bound state energies of same-charged pairs interacting via a repulsive potential within the gap. Notably when the binding energy corresponds to the middle of the gap, excitons and electron-electron pairs form zero energy states, which are currently a focus of research for a broad quantum computing community.

On first inspection is seems natural to solve for the simplest of potentials, the square well - defined by an abrupt step. However, the derivative of the potential results in Dirac delta functions, centered at the potential's walls, entering into Eq. (\ref{GrindEQ2}). Indeed, if the spinor components are a function of the potential, then any piecewise potential and its derivative should be continuous throughout the whole of space to assure that all the spinor components and their derivatives are also continuous. Therefore, it is natural to either solve for a truly smooth and continuous potential \cite{hartmann2010smooth,hartmann2014quasi} or a piecewise potential which has a smooth derivative throughout all of space.
\begin{figure}
\includegraphics[width=8cm,angle =270]{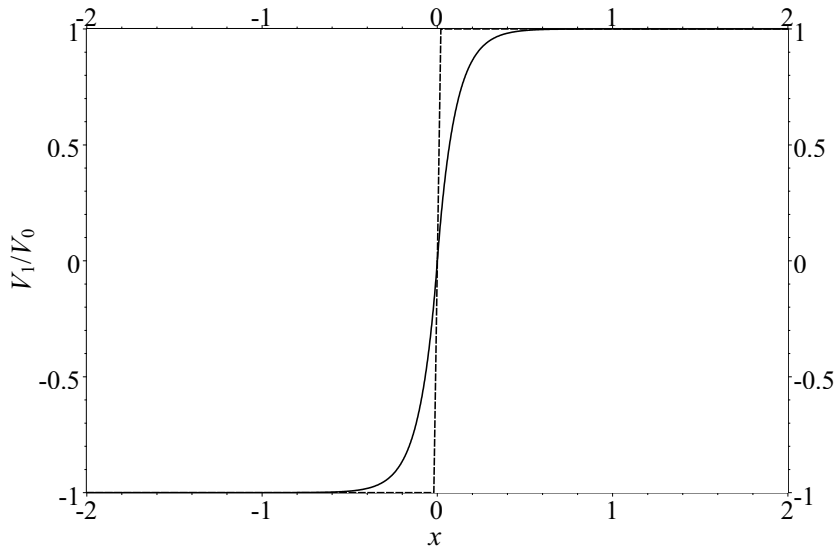}
\caption{
The ``smooth" Klein step for $L = 0.1$  (solid line) and $L = 0.001$ (dashed line) respectively
}
\label{fig:step}
\end{figure}
A smooth step potential can be defined as:
\begin{equation}
V_{1}=\left\{ \begin{array}{lll}
V_{0}\left[1-\exp\left(-z\right)\right] & , & z>0\\
V_{0}\left[\exp\left(z\right)-1\right] & , & z<0
\end{array}\right.
\label{eq:pot_step}
\end{equation}
and $V_{0} $ is half the height of the step. This potential belongs to a class of quantum models which are quasi-exactly solvable \cite{turbiner1988quantum,ushveridze1994quasi,bender1998quasi,downing2013solution,hartmann2014bound,hartmann2014quasi} and it shall be shown that in the limit that the potential transforms into the Heaviside step function the wavefunctions can be expressed in terms of elementary functions. In Fig. (\ref{fig:step}) we plot the smooth step potential for $L = 0.1$ and $0.001$. In the limit that $L \mathrm{\to} 0$ the potential tends towards the Heaviside step function and the potential becomes truly flat at the origin. It is convenient to consider one step only and take into account the second wall of the well by imposing the following wave function symmetry conditions at $z=-W/(2L)$:
\begin{equation}
\psi _{3} \left(-\frac{W}{2L} \right)=0,
\end{equation}
\begin{equation}
\left.\frac{\partial\psi_{3}}{\partial z}\right|_{-\frac{W}{2L}}=0,
\label{GrindEQ3}
\end{equation}
which correspond to odd and even modes of the square well respectively. $W$ is defined by the spatial extension of the interaction and the depth, $2V_{0} $, is obtained from the electrostatic attraction between the quasi particles. Since the two particles lie on the surface of the nanotube, the attractive interaction potential vanishing at infinity takes the form $U\left(x\right)\approx -e^{2} /(\varepsilon \sqrt{x^{2} +d^{2} } ).$ Here $\varepsilon$ is the effective dielectric constant and $d$ is the short-range cut-off parameter, which is of the order of the nanotube diameter. Therefore a realistic potential should be non-divergent, thus making the square well potential $V(x)=V_1(x)-V_0$ with $V_1(x)$ given by Eq.~\eqref{eq:pot_step} a good first order approximation for the short-range exciton interaction.

For $z<0$ the solution of Eq.~\eqref{GrindEQ2} is found to be
\begin{equation}
\psi_{3,\,\mathrm{I}}=\sum_{s_{\beta}}
A_{s_{\beta}}{\rm H_{c}}\left(\alpha_{+},\beta_{+},-2,\alpha_{+}^{2}/2,1-\alpha_{+}^{2}/2;Z_{+}\right)Z_{+}^{\frac{1}{2}\beta_{+}}
\exp\left(\frac{1}{2}\alpha_{+}Z_{+}\right),
\label{GrindEQ4}
\end{equation}
where $\alpha_{+}=s_{\alpha} i\left(V_{0} +\widetilde{E}\right)$, $\beta_{+} =s_{\beta} \sqrt{4\Delta ^{2} -\left(V_{0} +\widetilde{E}\right)^{2} }$, $s_{\alpha}=\pm1$, $s_{\beta}=\pm1$, $Z_{+} =V_{0} \exp \left(z\right)/(V_{0} +\widetilde{E})$ and $A_{s_{\beta}}$ are constants. Here we introduce $\widetilde{E}=E+V_0$. ${\rm H_{c}}\left(\alpha,\beta,\gamma, \delta,\eta  ; Z_{+}\right)$ is the Frobenius solution to the Heun confluent equation, which has two regular singularities at $Z_{+} =0$ and $1$ and one irregular singularity located at $\infty$ \cite{ronveaux1995heun}. The power series is computed about the origin and diverges at $Z_{+} =1$. It should be noted that for complex $\beta_{+}$ exchanging the sign of $\alpha_{+}$ results in the same eq.~(\ref{GrindEQ4}).
An analytic continuation of the power series can be obtained by expanding the solution about the second regular singularity $Z_{+} =1$ and matching the two series and their derivatives in between the singularities. The second pair of independent solutions can be constructed about the point $Z_{+} =1$ via the identity\cite{maier2007192}:
\begin{equation}
{\rm H_{c}}\left(\alpha ,\beta ,\gamma , \delta  ,\eta ; Z_{+}\right)
=
G_{1}
{\rm H_{c}}\left(-\alpha , \gamma , \beta , -\delta  ,\eta +\delta ; 1-Z_{+}\right)
+G_{2}
\left(-1+Z_{+}\right)^{-\gamma }
{\rm H_{c}}\left(-\alpha , -\gamma , \beta , -\delta , \eta +\delta  ; 1-Z_{+}\right).
\end{equation}
However ${\rm H_{c}}\left(-\alpha _{+} ,-2, \beta_{+} , -\alpha _{+}^{2} /2,1;1-Z_{+} \right)$ diverges since $\gamma =-2$ unless $\alpha _{+}^{2} =\beta _{+}^{2} $ i.e. for $\Delta =0$. Therefore, for non-zero $\Delta$, $G_{1}=0$ and the solution to Eq.~(\ref{GrindEQ2}) can be written as:
\begin{equation}
\psi_{3,\,i}=
\left(1-Z_{+}\right)^{2}\sum_{s_{\alpha},\,s_{\beta}}
C_{s_{\alpha},s_{\beta}}{\rm H_{c}}\left(-\alpha_{+},2,\beta_{+},-\alpha_{+}^{2}/2,1;1-Z_{+}\right)Z_{+}^{\frac{1}{2}\beta_{+}}
\exp\left(\frac{1}{2}\alpha_{+}Z_{+}\right),
\label{GrindEQ5}
\end{equation}
where $C_{s_{\alpha},s_{\beta}}$ are constants. However,
${\rm H_{c}}\left(-\alpha _{+} ,2,\beta_{+} ,-\alpha _{+}^{2} /2,1;1-Z_{+} \right)
Z_{+}^{\frac{1}{2}\beta_{+}}={\rm H_{c}}\left(-\alpha _{+} ,2,-\beta _{+} ,-\alpha _{+}^{2} /2,1;1-Z_{+} \right)
Z_{+}^{-\frac{1}{2}\beta_{+}}$ therefore we may set $c_{s_{\alpha},-1}=0$.

For $z>0$ we obtain the solution
\begin{equation}
\psi_{3,\,\mathrm{II}}=\sum_{s_{\beta}}
B_{s_{\beta}}{\rm H_{c}}\left(\alpha_{-},\beta_{-},-2,\frac{1}{2}\alpha_{-}^{2},1-\frac{1}{2}\alpha_{-}^{2},Z_{-}\right)Z_{-}^{\frac{1}{2}\beta_{-}}
\exp\left(\frac{1}{2}\alpha_{-}Z_{-}\right),
\label{GrindEQ6}
\end{equation}
where $\alpha _{-} =s_{\alpha} i\left(V_{0} -\widetilde{E}\right)$, $\beta _{-} =s_{\beta} \sqrt{4\Delta ^{2} -\left(V_{0} -\widetilde{E}\right)^{2} } $, $Z_{-} =V_{0} \exp \left(-z\right)/(V_{0} -\widetilde{E})$ and the expansion about $Z_{-} =1$ is given by
\begin{equation}
\psi_{3,\,\mathrm{ii}}=\left(1-Z_{-}\right)^{2}\sum_{s_{\alpha}}
D_{s_{\alpha}}{\rm H_{c}}\left(-\alpha_{-},2,\beta_{-},-\alpha_{-}^{2}/2,1;1-Z_{-}\right)Z_{-}^{\frac{1}{2}\beta_{-}}
\exp\left(\frac{1}{2}\alpha_{-}Z_{-}\right),
\label{GrindEQ7}
\end{equation}
where $D_{s_{\alpha}}$ are constants. It is clear from Eq. (\ref{GrindEQ6}) and (\ref{GrindEQ7}) that for the function to decay at infinity we require $4\Delta ^{2} >\left(V_{0} -\widetilde{E}\right)^{2} $, therefore $D_{-1} =0$. A real $\beta _{-} $ means that all bound states of positive energy lie within the band gap of the two-body system.

In what follows we shall restrict ourselves to analyzing bound states whose energy is above the centre of the band gap i.e. $\widetilde{E}-V_{0}>0$. Paired states of negative energy shall be a topic of future study. For photo-created electron-hole pairs the energy range $\widetilde{E}-V_{0}>0$ is sufficient. Indeed, variational calculations \cite{Pedersen20041007} of the binding energy in semiconductor nanotubes supported by experimental data give a value for the exciton binding energy of approximately $30\%$ of the band gap.
For the Frobenius solutions to converge we require that their arguments be less than $1$.
At the boundary $Z_{+}(0) =V_{0}/(V_{0} +\widetilde{E})$ and $Z_{-}(0) =V_{0}/(V_{0} - \widetilde{E})$. Therefore, $0<Z_{+}(0)\leq 1/2$ whereas $Z_{-}(0)<0$. Hence, for $z>0$, we restrict ourselves to the Frobenius solutions of argument $Z_{-}$ whereas for $z<0$, the Frobenius solutions of arguments $Z_{+}$ and $1-Z_{+}$ are valid at the boundary $z=0$. However, the Frobenius solution of argument $1-Z_{+}$ diverges as $z\rightarrow0$ which occurs rapidly for $\left|z\right|>L$ as $L$ tends towards zero.

\begin{figure}
\includegraphics*[width=8cm,angle=-90]{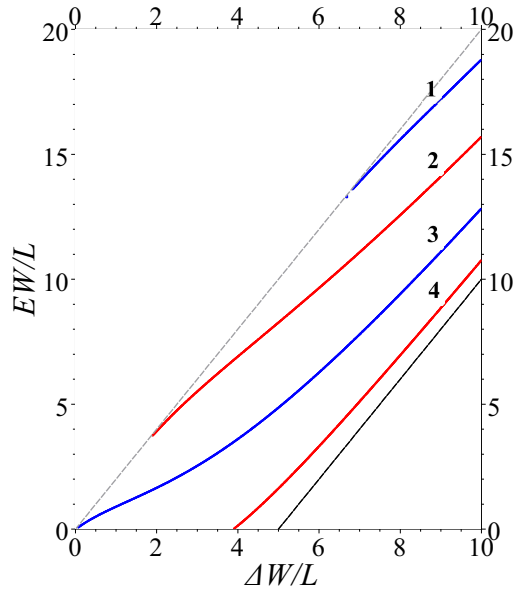}
\caption{(Color online) The energy spectrum of an electron-hole pair interacting via a one-dimensional smooth square well as a function of $\Delta W/L$ for $2 W V_{0} /L=10$. The gray-dashed (topmost line) and black-dashed (bottommost line) lines represent $E= 2 \Delta$  and $E -2V_{0} = 2 \Delta$  respectively. Lines $2$ and $4$ (in red) correspond to the even modes, while lines $1$ and $3$ (in blue) correspond to the odd modes.}
\label{fig:spectrum}
\end{figure}

\begin{figure}
\includegraphics[height=8cm,angle=270]{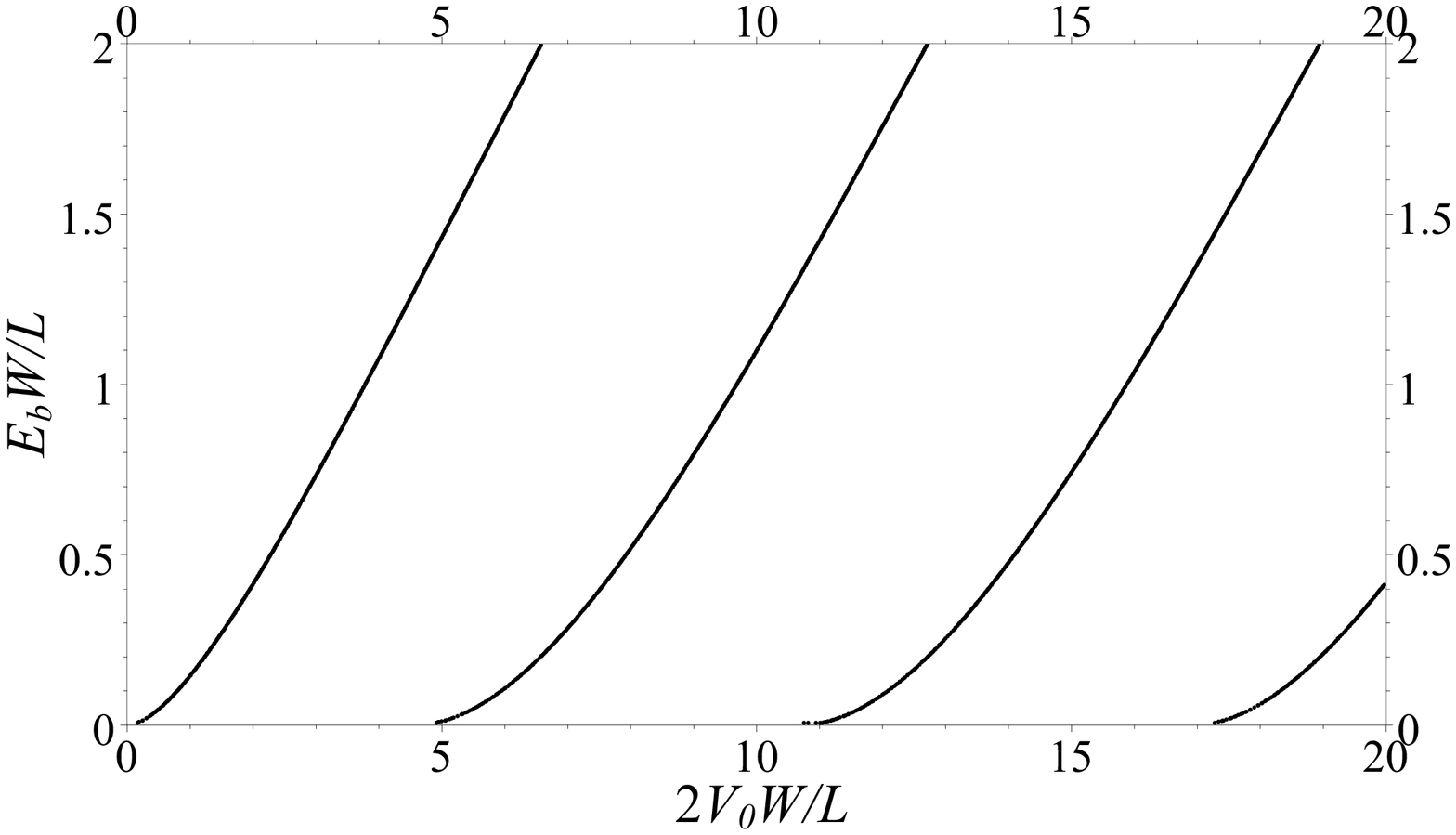}
\includegraphics[height=8cm,angle=270]{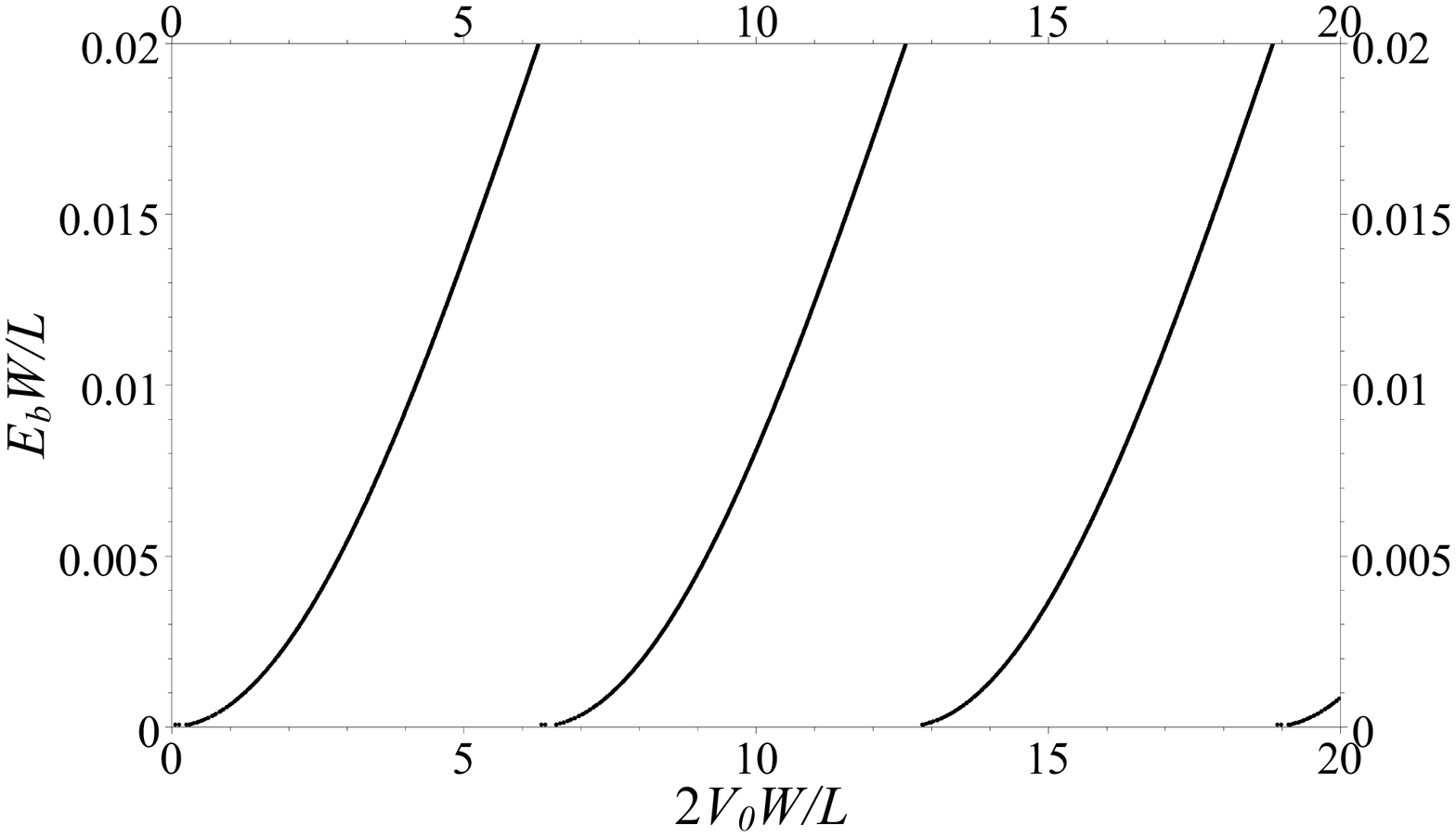}
\caption{
The dependence of the exciton energy $E_{b}$ on the interaction strength $2V_0$: The left-hand side is for a semiconductor nanotube with $\Delta W/L=1$; The right-hand side is for a narrow-gap tube with $\Delta W/L=0.01$. The different lines correspond to different excitonic states.
}
\label{Fig_potential strength_small}
\end{figure}

At the boundary the Frobenius solutions about $Z_{\pm }=0$ may be expanded as a power series in $\alpha _{\pm }$ and $\beta _{\pm }$. By considering first-order powers of $L$ only, one may write
\begin{equation}
{\rm H_{c}}\left(\alpha_{\pm} ,\beta _{\pm } , -2, \alpha _{\pm }^{2} /2,1-\alpha _{\pm }^{2} /2; Z_{\pm } \left(0\right)\right)
\exp \left(\frac{1}{2} \alpha_{\pm } Z_{\pm} \left(0\right)\right)
\left[Z_{\pm } \left(0\right)\right]^{\frac{1}{2} \beta _{\pm } } \approx 1
+
\frac{1}{2} \beta_{\pm } \left[\ln \left(Z_{\pm } \left(0\right)\right)-Z_{\pm } \left(0\right)\right]
\label{approx}
\end{equation}
and in the limit that $z\to \infty $, $Z_{+} \to 0$ and the Heun function of argument zero has a value of unity therefore which allows the asymptotic wave function to be written as
\begin{equation}
\mathop{\lim }\limits_{L\to 0} \left( \psi_{3,\,\mathrm{I}} \right)=
A_{1} \exp \left(\frac{1 }{2} \beta_{+} z\right)+A_{-1} \exp \left(-\frac{1 }{2} \beta_{+} z\right).
\label{GrindEQ9}
\end{equation}
Using the approximation Eq.~(\ref{approx}) and equating Eq.~(\ref{GrindEQ4}) and (\ref{GrindEQ5}) and their derivatives at $z=0$ allows the equation \eqref{GrindEQ9} to be written as
\begin{equation}
\lim\limits _{L\to0}\left(  \psi_{3,\,\mathrm{I}}   \right)= B_{1}\left[\cos\left(\frac{z}{2}\sqrt{\left(V_{0}+\widetilde{E}\right)^{2}-4\Delta^{2}}\right)+\frac{V_{0}+\widetilde{E}}{V_{0}-\widetilde{E}}\frac{\sqrt{4\Delta^{2}-\left(V_{0}-\widetilde{E}\right)^{2}}}{\sqrt{\left(V_{0}+\widetilde{E}\right)^{2}-4\Delta^{2}}}\sin\left(\frac{z}{2}\sqrt{\left(V_{0}+\widetilde{E}\right)^{2}-4\Delta^{2}}\right)\right],
\label{Asym_one}
\end{equation}
and for $V_{0}$=0, Eq.~(\ref{Asym_one}) reduces to the plane wave of the two-bodied wave function of wave-vector $\sqrt{\widetilde{E}^{2}-4\Delta^{2}}/2L$. The other components are obtained via the relationships given by Eq.~(\ref{GrindEQ3}). No further matching conditions are required since $\psi_{3}$, the potential and their derivatives are matched at the boundary (note this is not the case if the derivative of the potential is discontinuous). For bound states,  $\psi_{3}$ exponentially decays outside of the well, therefore all the other spinor components will decay too,  and since $\widetilde{E}>V_0$ there will be no singularities in the other spinor components. The odd modes of the square well are therefore given by
\begin{equation}
\frac{V_{0} -\widetilde{E}}{V_{0} +\widetilde{E}} \frac{\sqrt{\left(V_{0} +\widetilde{E}\right)^{2} -4\Delta ^{2} } }{\sqrt{4\Delta ^{2} -\left(V_{0} -\widetilde{E}\right)^{2} } } -\tan \left(\frac{W}{4L} \sqrt{\left(V_{0} +\widetilde{E}\right)^{2} -4\Delta ^{2} } \right)=0 ,
\label{odd_Pos}
\end{equation}
and the even modes are given by
\begin{equation}
\frac{V_{0} -\widetilde{E}}{V_{0} +\widetilde{E}} \frac{\sqrt{\left(V_{0} +\widetilde{E}\right)^{2} -4\Delta ^{2} } }{\sqrt{4\Delta ^{2} -\left(V_{0} -\widetilde{E}\right)^{2} } } +\cot \left(\frac{W}{4L} \sqrt{\left(V_{0} +\widetilde{E}\right)^{2} -4\Delta ^{2} } \right)=0.
\label{even_Pos}
\end{equation}
The two transcendental equations can be solved graphically or via other standard root-finding methods. In Fig.~(\ref{fig:spectrum}) we plot the obtained energy spectrum for $V_{0}W/L=10$ and in Fig. (\ref{Fig_potential strength_small}) we show the dependence of the static exciton energy on the depth of the well for two different values: $\Delta W/L=1$ and $\Delta W/L=0.01$ corresponding to the cases of semiconductor and narrow-gap nanotubes respectively. For both narrow-gap and semi-conducting tubes, when $V_{0}W/L$ is small, there is only one bound state. As $V_{0}$ increases, the binding energy, defined as $E_{b}=2\Delta - \widetilde{E}$, increases until it is equal to the value of the band gap, upon which the exciton enters the continuum of states and disassociates.

As mentioned in the introduction, the question of exactly how excitonic effects influence the optical processes in narrow-gap nanotube is important for prospective THz devices. Previous works suggest that the one-dimensional Van Hove singularity is suppressed by excitonic effects for both long-range and short-range electron-hole interaction potentials \cite{portnoi2013excitons} and prominent peaks arise in the absorption spectrum which coincide with the exciton bound state energies. Our analytic results are therefore extremely important in determining the role of excitonic effects, since knowledge of the eigenvalues and functions at zero separation allows one, in principle, to calculate the absorption coefficient via the Elliot formula. This shall be a topic of future study.

Unlike for the Gauss hypergeometric series, for the Heun functions the general formulae connecting solutions about two different singular points for arbitrary parameters is not known. To analyze the behavior of Eq.~(\ref{GrindEQ7}) we make use of the following approximation. In the limit that $L\to 0$
%
\begin{equation}
{\rm H_{c}}\left(-\alpha_{\pm} , 2, \beta_{\pm} ,0,1; 1-Z_{\pm} \right) \approx  {}_{2}F_{1} \left(2+\frac{1}{2} \beta_{\pm} , 1+\frac{1}{2} \beta_{\pm} ; 3; 1-Z_{\pm} \right)
-\alpha_{\pm} \left[\frac{1-Z_{\pm} +\ln \left(Z_{\pm} \right)}{1-Z_{\pm} } \right],
\label{Heuntohyper}
\end{equation}
where $_{2}F_{1}$ is the Gauss hypergeometric function. This enables Eq.(\ref{GrindEQ5}), at the well's edge to be written as
\begin{equation}
\psi_{3,\,i}\approx-\sum_{s_{\alpha}}C_{s_{\alpha},1}\left(2+\alpha_{+}\right)\left[\left(1-Z_{+}\right)+\ln\left(Z_{+}\right)\right]
\label{GrindEQ5_Prox}
\end{equation}
and as $Z_{+} \rightarrow 0$ to be be expressed as
\begin{equation}
\psi_{3,\,i}\approx-\sum_{s_{\alpha}}C_{s_{\alpha},1}\left(2+\alpha_{+}\right)\left[\cosh\left(\frac{1}{2}\beta_{+}z\right)+\frac{2}{\beta_{+}}\sinh\left(\frac{1}{2}\beta_{+}z\right)\right].
\label{farfield_L_Shift}
\end{equation}
For arbitrary $\widetilde{E}$ and $V_0$ it is not possible to assure the continuity of  Eq.~(\ref{GrindEQ5_Prox}) and Eq.~(\ref{GrindEQ6}) and their derivatives at the step-edge using the approximate wave functions. However, both Frobenius solutions of arguments $Z_{+}$ and $1-Z_{+}$ are valid at the boundary $z=0$. Therefore, we may search for solutions which are a superposition of both Frobenius solutions inside the well. Upon substituting the approximate wave functions Eq. (\ref{approx}) and Eq. (\ref{GrindEQ5_Prox}) into the boundary conditions:
\begin{equation}
\psi_{3,\,\mathrm{II}}\left(0\right)=\psi_{3,\,\mathrm{I}}\left(0\right)+\psi_{3,\,i}\left(0\right),
\quad
\left.\frac{\partial\psi_{3,\,\mathrm{II}}}{\partial z}\right|_{0}=\left.\frac{\partial\psi_{3,\,\mathrm{I}}+\psi_{3,\,i}}{\partial z}\right|_{0}.
\end{equation}
and making use of the asymptotic expressions Eq. (\ref{GrindEQ9}) and Eq.(\ref{farfield_L_Shift}) we find that if $A_{1}$ or $A_{-1}$ are zero then we restore the eigenvalues obtained in Eq. (\ref{odd_Pos}) and Eq. (\ref{even_Pos}). For the case of non-zero $A_{1}$ and $A_{-1}$ we find that:
\begin{equation}
A_{\pm1}\approx\frac{1}{2}\left[1\mp\frac{\beta_{-}}{\beta_{+}}\frac{1-Z_{-}}{1-Z_{+}}\right]B_{1}\mp\frac{\omega}{\beta_{+}},
\end{equation}
where $\omega=-\sum_{s_{\alpha}}C_{s_{\alpha},1}\left(2+\alpha_{+}\right)$, which allows the superposition of $\psi_{3,\,\mathrm{I}}\left(-W/2\right)+\psi_{3,\,i}\left(-W/2\right)$ to be expressed as
\begin{equation}
B_{1}\left[\cosh\left(\frac{1}{2}\beta_{+}z\right)-\frac{\beta_{-}}{\beta_{+}}\frac{1-Z_{-}}{1-Z_{+}}\sinh\left(\frac{1}{2}\beta_{+}z\right)\right]+\omega\cosh\left(\frac{1}{2}\beta_{+}z\right).
\end{equation}
Therefore, for the case of finite, non-divergent $A_{1}$ and $A_{-1}$, we require that $\omega \propto L$ and in this instance the eigenvalues are restored.

\section{The Moving Exciton}
The spinor components of the eigenfunctions of a square well need
not necessarily be continuous at the well's edge \cite{alberto1996relativistic}
since they are solutions to a system of first order differential equations
containing a potential which is itself discontinuous. By analyzing the behavior
of the wavefunction for our smooth potential, in the limit in which it approaches a smooth square well, one can obtain the appropriate boundary conditions for the spinor components of the true square well at the well's edge.
For our model potential as $L\rightarrow0$ the wavefunction remain continuous at the well's edge, however, the derivative of the spinor components may tend to infinity which corresponds to an abrupt jump in a square well's
wavefunction i.e. a discontinuity. 
The derivatives of the spinor components of the smooth square well are given by the expressions:
\begin{equation}
\frac{\partial\psi_{1}}{\partial x}=\frac{1}{L}\left[\pm\frac{1}{\left(1-Z_{\pm}\right)}\psi_{1}-\Delta\psi_{4}\right],
\nonumber
\end{equation}
\begin{equation}
\frac{\partial\psi_{4}}{\partial x}=-\frac{1}{2}\left[\frac{\left(E-V\right)^{2}-4\Delta^{2}}{\left(E-V\right)L}\right]\psi_{3},
\nonumber
\end{equation}
\begin{equation}
\frac{\partial\psi_{3}}{\partial x}=\pm\frac{1}{L}\frac{\partial\psi_{3}}{\partial Z_{\pm}}.
\nonumber
\end{equation}
From Eq.~(\ref{approx}) $\left.\partial\psi_{3}/\partial Z_{\pm}\right|_{z=0}\propto L$, hence it can be seen from the above expressions that when evaluated at the
well's edge $\partial\psi_{1}/\partial x$ diverges as $L\rightarrow0$, while all the other components and their derivatives remain finite. Away from the well's edge, all the spinor components of positive energy bound states are non-divergent. Therefore when solving the same problem for an abrupt step potential, and analyzing to the left and
to the right of the well's wall, the wavefunctions to be matched are $\psi_{3}$ and $\psi_{4}$ and not $\psi_{1}$.
However, though $\psi_{3}$ and $\psi_{4}$ are continuous across the square well, their derivatives are not since they are functions of the potential derivative.

\begin{figure}
\includegraphics[height=12cm,angle=270]{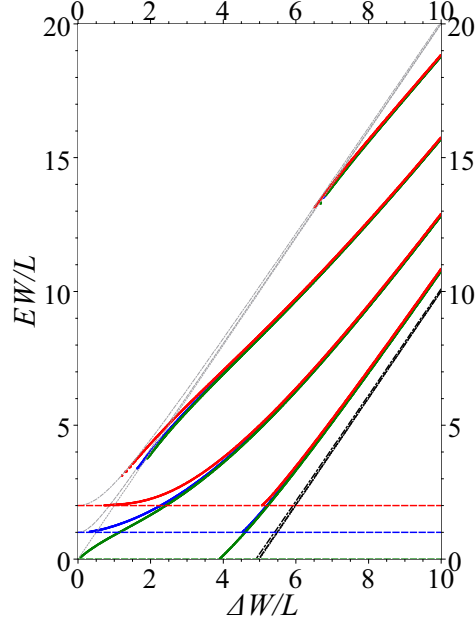}
\caption{
(Color online) The thick gray dotted, dashed and dotted-dashed lines denote $E=\sqrt{\widetilde{K}^{2}+4\Delta^{2}}$ for $\widetilde{K}=0,1$ and $2$ respectively, while their black counterparts denote $E=\sqrt{\widetilde{K}^{2}+4\Delta^{2}}-V_0$. The solid green, blue and red lines (the lowest, second from the bottom and top line in each set of lines respectively) correspond to the eigenvalues of $\widetilde{K}=0,1$ and $2$ respectively, and their dashed counterparts mark $E=\widetilde{K}$.
}
\label{Spectrum_K}
\end{figure}

\begin{figure}
\includegraphics[height=10cm,angle=270]{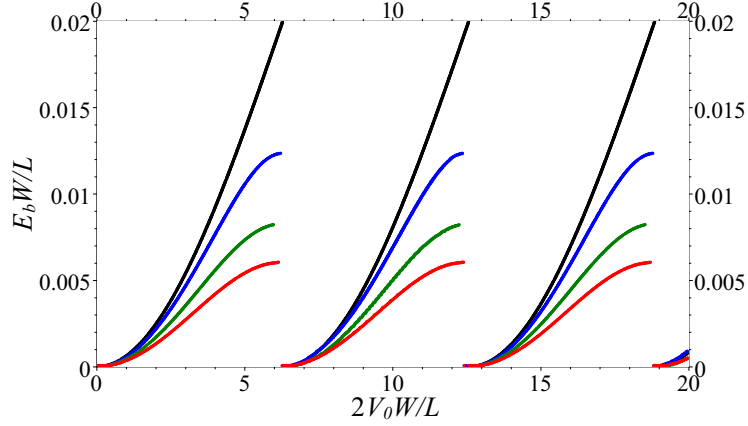}
\caption{
(Color online) The dependence of the exciton binding energy, $E_b$, on the interaction strength $2V_0$ with $\Delta W/L=0.01$ for $K=0$ (topmost set of lines, in black), $K=\Delta$ (second from top set of lines, in blue), $K=2\Delta$ (third from top set of lines, in green) and $K=3\Delta$ (bottommost set of lines, in red).
}
\label{Fig_potential strength_K}
\end{figure}

For the case of the square well, of depth $\widetilde{V}_{0}$ and width $W$, centered about the origin,
Eq.~(\ref{GrindEQ2}) becomes
\begin{equation}
\frac{\partial^{2}\psi_{3}}{\partial x^{2}}+\lambda\psi_{3}=0,
\end{equation}
where $L^{2} \lambda=\left[E^{2}-4\Delta^{2}\right]/4$ inside the well
and $L^{2} \lambda=\left[\left(E+\widetilde{V}_{0}\right)^{2}-4\Delta^{2}\right]/4$ outside of the well, which admits the solution
\begin{equation}
\psi_{3}=\left\{ \begin{array}{lll}
A_{\mathrm{I}}\cos\left(\frac{\sqrt{\left(E+\widetilde{V}_{0}\right)^{2}-4\Delta^{2}}}{2L}x\right)+A_{\mathrm{II}}\sin\left(\frac{\sqrt{\left(E+\widetilde{V}_{0}\right)^{2}-4\Delta^{2}}}{2L}x\right) &,& x<\left|\frac{W}{2}\right|\\
\pm B\exp\left(-\frac{\sqrt{4\Delta^{2}-E^{2}}}{2L}\left|x\right|\right) &,& x>\left|\frac{W}{2}\right|
\end{array}\right.
\end{equation}
where $A_{\mathrm{I}}$, $A_{\mathrm{II}}$ and $B$ are constants and for odd modes $A_{\mathrm{I}}=0$ while for even $A_{\mathrm{II}}=0$.
Using the continuity of the function $\psi_{3}$ and $\psi_{4}$ at the well's edge one restores the result obtained for the smooth square well. However, unlike the smooth square well one may obtain exact solutions with finite $K$ for the true square well. 
For the case of an exciton possessing
finite total momentum along the nanotube axis $\psi_{2}$ cannot be eliminated, and both $\partial\psi_{1}/\partial x$ and $\partial\psi_{2}/\partial x$
diverge at the well's edge. By matching $\psi_{3}$ and $\psi_{4}$ at the well's edge one obtains the following eigenvalue relations:
\begin{equation}
\tan\left(\frac{W}{4L}\sqrt{\frac{\left(E+\widetilde{V}_{0}\right)^{2}-\widetilde{K}^{2}-4\Delta^{2}}{\left(E+\widetilde{V}_{0}\right)^{2}-\widetilde{K}^{2}}\left(E+\widetilde{V}_{0}\right)^{2}}\right)+\frac{E}{\left(E+\widetilde{V}_{0}\right)}\sqrt{\frac{\left(E^{2}-\widetilde{K}^{2}\right)\left(E+\widetilde{V}_{0}\right)^{2}\left[\left(E+\widetilde{V}_{0}\right)^{2}-\widetilde{K}^{2}-4\Delta^{2}\right]}{E^{2}\left[\left(E+\widetilde{V}_{0}\right)^{2}-\widetilde{K}^{2}\right]\left[4\Delta^{2}-\left(E^{2}-\widetilde{K}^{2}\right)\right]}}=0,
\label{eqn:K_spectrum_1}
\end{equation}
\begin{equation}
\cot\left(\frac{W}{4L}\sqrt{\frac{\left(E+\widetilde{V}_{0}\right)^{2}-\widetilde{K}^{2}-4\Delta^{2}}{\left(E+\widetilde{V}_{0}\right)^{2}-\widetilde{K}^{2}}\left(E+\widetilde{V}_{0}\right)^{2}}\right)-\frac{E}{\left(E+\widetilde{V}_{0}\right)}\sqrt{\frac{\left(E^{2}-\widetilde{K}^{2}\right)\left(E+\widetilde{V}_{0}\right)^{2}\left[\left(E+\widetilde{V}_{0}\right)^{2}-\widetilde{K}^{2}-4\Delta^{2}\right]}{E^{2}\left[\left(E+\widetilde{V}_{0}\right)^{2}-\widetilde{K}^{2}\right]\left[4\Delta^{2}-\left(E^{2}-\widetilde{K}^{2}\right)\right]}}=0,
\label{eqn:K_spectrum_2}
\end{equation}
where $\widetilde{K}=KL$ and we restrict ourselves to the case where $\left(E+\widetilde{V}_{0}\right)^{2}-\widetilde{K}^{2}>0$ to assure there are no singularities in the spinor components and ensure that $4\Delta^{2}-\left(E^{2}-\widetilde{K}^{2}\right)>0$ and $E^{2}-\widetilde{K}^{2}>0$ for the states to decay outside of the well. Here Eq. (\ref{eqn:K_spectrum_1}) is for odd $\psi_{3}$
and Eq. (\ref{eqn:K_spectrum_2}) is for even $\psi_{3}$. It should be noted that in the delta function limit considered in the next section, Eq. (\ref{eqn:K_spectrum_1}) and Eq. (\ref{eqn:K_spectrum_2}), give the same result as for the delta function of small strength. In general Eq.~(\ref{eqn:K_spectrum_1}) and Eq.~(\ref{eqn:K_spectrum_2}) can be solved graphically or via other standard root-finding methods. In Figure~(\ref{Spectrum_K}) we plot the obtained energy spectrum as a function of $\Delta W/L$ for $\widetilde{V}_{0}W/L=10$ for a range of $K$ values and it can be seen that as $K$ increases, the bands blue shift. In Figure (\ref{Fig_potential strength_K}) we show the dependence of the dynamic exciton energy on the depth of the well for a narrow-gap nanotube defined by $\Delta W/L=0.01$. The increase of $K$ naturally shifts the excitonic states to higher energies.

\section{The Delta function Potential}
Let us now consider the limit of very weak electron-hole attraction in the non-relativistic regime: $\left|2V_{0}\right|\ll\Delta$.
In this limit approximate solutions to Eq.~(\ref{odd_Pos}) and Eq.~(\ref{even_Pos}) can be obtained for small binding energies.  We find that for small binding energies $E_{b}=\Delta V_{0}^{2}W^{2}/L^2$. Therefore for a very narrow, deep well where 
$2 V_{0}/L \gg W$ with $2 V_{0}W/L=\alpha$ the binding energy is $E_{b}=\Delta \alpha^{2}/4$, thus we recover the non-relativistic solution for an attractive delta function potential of strength $\alpha$. The strength of the potential $\alpha$, i.e. $V=-\alpha L \delta (x) $, can be estimated as a product of the strength of the realistic potential and its width.

\begin{figure}
\includegraphics*[width=7cm,angle=-90]{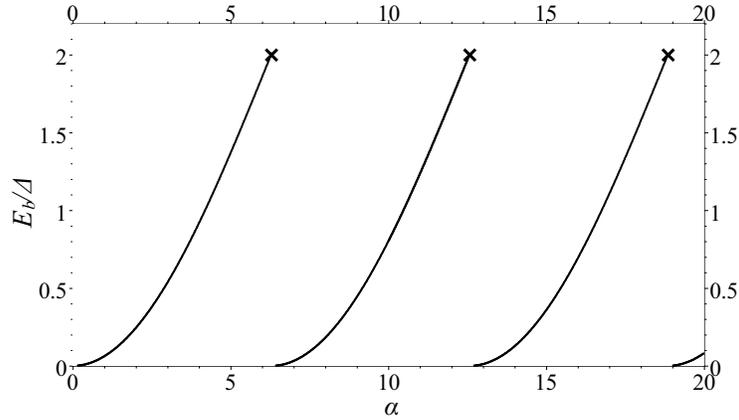}
\caption{
The dependence of the exciton binding energy, $E_{b}$, on the interaction strength $\alpha$, for an electron-hole pair interacting via the interaction potential, Eq.~(\ref{eq:pot_step}), in the delta function limit. The different lines correspond to different excitonic states. The crosses indicate when $E_{b}=2\Delta$, which occurs at $\alpha=2\left(1+n\right)\pi$, where $n=0,1,2,3$ etc.
}
\label{fig:Delta_Alpha}
\end{figure}

We shall now consider the case when the interaction potential, Eq.~(\ref{eq:pot_step}),
tends towards the Delta function potential $V_{1}(x)=-\alpha L\delta\left(x\right)+V_{0}$ in the non-classical regime. 
In the limit that $2V_{0}/L\rightarrow\infty$ and $W\rightarrow0$ such that $2V_{0}W/L=\alpha$ and $\Delta W/L\ll1$, Eq.~(\ref{odd_Pos}) and Eq.~(\ref{even_Pos}) admit the approximate solutions
\begin{equation}
E=-2\Delta\frac{\tan\left(\frac{\alpha}{4}\right)}{\sqrt{1+\tan^{2}\left(\frac{\alpha}{4}\right)}}
\label{delta_1}
\end{equation}
and
\begin{equation}
E=2\Delta\frac{\cot\left(\frac{\alpha}{4}\right)}{\sqrt{1+\cot^{2}\left(\frac{\alpha}{4}\right)}}
\label{delta_2}
\end{equation}
for odd and even modes respectively. For a given $\Delta$, if Eq.(\ref{delta_1}) is negative, Eq.(\ref{delta_2}) is positive and vice-versa. Therefore, the delta function can at most contain one bound state of positive energy. This can be seen in Fig. (\ref{fig:Delta_Alpha}), where the appearance of a higher order solutions coincides with the dissociation of the lower order one. In the limit of very weak electron-hole attraction we find that for small binding energies, $E_{b}\approx\Delta \alpha^2 /16$.

There are many other smooth analytic approximations to the delta function.
The hyperbolic secant potential, $V=-\alpha/\left[\pi \cosh\left(x/L\right)\right]$ is known to admit analytical expressions for zero energy when $\alpha=\left(1+2n\right)\pi+\sqrt{1+4\Delta^{2}}\pi$ \cite{hartmann2011excitons}.
In the delta function limit, i.e. as $L\rightarrow0$ the condition becomes
$\alpha=2\left(1+n\right)\pi$, where $n=0,1,2,\ldots$ Restricting ourselves to the modes of relevant parity, we find the same conditions hold true for Eq.~(\ref{delta_1}) and Eq.~(\ref{delta_2}) at zero energy.

It should be noted that for odd solutions of $\psi_{3}$, the delta function can be solved imposing that $\psi_{3}$ is discontinuous at the origin, whereas $\psi_{4}$ is continuous but its derivative is not. Let us consider the potential $V=-\alpha L\delta\left(x\right)$; expressing Eq. (\ref{GrindEQ1}) in terms of $\psi_{4}$ and integrating across the interval [$0_{-}$; $0_{+}$] yields
\begin{equation}
E\left(\left.\frac{\partial\psi_{4}}{\partial \widetilde{x}}\right|_{0_{+}}-\left.\frac{\partial\psi_{4}}{\partial \widetilde{x}}\right|_{0_{-}}\right)+\frac{1}{4}\alpha\left(E^{2}-4\Delta^{2}\right)\psi_{4}\left(0\right)=0,
\label{DTrue}
\end{equation}
where $\widetilde{x}=x/L$. A second relation can be found in the regions $x>0$ and $x<0$: $\frac{\partial^{2}\psi_{4}}{\partial\widetilde{x}^{2}}-\frac{1}{4}\left(4\Delta^{2}-E^{2}\right)\psi_{4}=0$, which admits the solution $\psi_{4}=G_{4}\exp\left(-\sqrt{4\Delta^{2}-E^{2}}\left|\widetilde{x}\right|/2\right)$. Substituting the defintion of $\psi_{4}$ into Eq. (\ref{DTrue}) results in the eigenvalue $E=-2\Delta\alpha/\sqrt{16+\alpha^{2}}$, which for small values of $\alpha$ (when $\tan\left(\alpha/4\right)\approx\alpha/4$) is in agreement with Eq.(\ref{delta_1}). Since $\psi_{4}$ is even, it follows that $\psi_{1}$ and $\psi_{3}$ are odd.
For even $\psi_{3}$ we must solve Eq. (\ref{GrindEQ2}). Integrating Eq. (\ref{GrindEQ2}) across the interval [$0_{-}$; $0_{+}$] yields
\begin{equation}
\frac{1}{E}\left(\left.\frac{\partial\psi_{3}}{\partial \widetilde{x}}\right|_{0_{+}}-\left.\frac{\partial\psi_{3}}{\partial \widetilde{x}}\right|_{0_{-}}\right)+\frac{1}{4}\alpha\psi_{3}\left(0\right)-\intop_{0_{-}}^{0_{+}}\frac{\Delta^{2}}{\left(E-V\right)}\psi_{3}d\widetilde{x}=0,
\label{eq:Even_3}
\end{equation}
and in the regions $x>0$ and $x<0$ we find that $\psi_{3}=G_{3}\exp\left(-\sqrt{4\Delta^{2}-E^{2}}\left|\widetilde{x}\right|/2\right)$. After regularization, the integral containing the delta function becomes zero, and the resulting eigenvalue is found to be $E=8\Delta/\sqrt{\alpha^{2}+16}$ which in the limit of small $\alpha$ agrees with Eq.(\ref{delta_2}).


We shall now consider the case of an exciton possessing finite total momentum along the nanotube axis formed by an electron-hole pair interacting via a delta-function potential. Repeating the same procedure for finite $K$, we find for even $\psi_{4}$, $\psi_{4}=G_{4}\exp\left(-\frac{1}{2}\sqrt{\frac{4\Delta^{2}+\widetilde{K}^{2}-E^{2}}{E^{2}-\widetilde{K}^{2}}}\left|E\widetilde{x}\right|\right)$ and
\begin{equation}
E=\pm\sqrt{\widetilde{K}^{2}+\frac{4\alpha^{2}\Delta^{2}}{16+\alpha^{2}}},
\label{delta_energy_1}
\end{equation}
where $\widetilde{K}=KL$. We also require that $4\Delta^{2}+\widetilde{K}^{2}>E^{2}$ to ensure the state is bound. For an even $\psi_{3}$, we integrate Eq. (\ref{GrindEQ2}) across the interval [$0_{-}$; $0_{+}$] to obtain
\begin{equation}
\frac{1}{E}\left(\left.\frac{\partial\psi_{3}}{\partial \widetilde{x}}\right|_{0_{+}}-\left.\frac{\partial\psi_{3}}{\partial \widetilde{x}}\right|_{0_{-}}\right)+\frac{1}{4}\alpha\psi_{3}\left(0\right)+
\intop_{0_{-}}^{0_{+}}\frac{\left(E-V\right)\Delta^{2}}{\widetilde{K}^{2}-\left(E-V\right)^{2}}\psi_{3}d\widetilde{x}=0.
\end{equation}
Regularizing the delta function requires that the denominator be non-singular, this imposes the requirement that $\widetilde{K}^{2}-\left(E-V\right)^{2}>0$. Since $\psi_{3}=G_{3}\exp\left(-\frac{1}{2}\sqrt{\frac{4\Delta^{2}+K^{2}-E^{2}}{E^{2}-K^{2}}}\left|E\widetilde{x}\right|\right)$ the integral containing the delta function becomes zero and the resulting eigenvalue is found to be
\begin{equation}
E=\pm\sqrt{\widetilde{K}^{2}+\frac{64\Delta^{2}}{16+\alpha^{2}}}.
\label{delta_energy_2}
\end{equation}
Comparing the rest frame to the moving frame, and treating the transverse momentum as an effective mass, one restores the energy-momentum relation given by special relativity.


\section{Conclusions}
Solutions were obtained in the rest frame for the quasi-one-dimensional two-body Dirac problem, for a smooth step interaction potential, in terms of Heun confluent functions. By symmetry, this potential was used to obtain the eigenvalues of a smooth square well,  which are found by solving a set of transcendental equations. Such a potential can be used to approximate the interaction potential between an electron-hole pair in a narrow-gap nanotube. The binding energy of these pairs, was found to never exceed the band gap and therefore at room temperature the electron-hole pairs should be fully ionized. Hence, undesirable effects due to dark excitons should not dominate optical processes in narrow-gap nanotubes. By analyzing the smooth square well's stationary excitonic wave functions, the appropriate boundary conditions were obtained for an abrupt square well which in turn enables the dynamic exciton energy levels to be found. We also consider delta function interaction - a highly non-trivial problem for relativistic particles - and show that different approximations for the delta function give the same result. Many of our results can be generalized for a pair of particles of the same charge, which as we show can have bound states within the gap.

The question of exactly how excitonic effects influence the optical processes in narrow-gap nanotube is still an outstanding problem. This piecewise potential not only serves as an important tool for analyzing the excitonic energy levels in a narrow gap nanotube, but the simplicity of the asymptotic forms of the wave function in the square limit coupled with their easily determinable eigenvalues, are extremely useful in determining the optical absorption spectra due to excitons in narrow-gap carbon nanotubes \cite{portnoi2013excitons}. It should also be noted that the Hamiltonian used in this paper is of the same form as certain types of graphene nanoribbons or armchair carbon nanotubes subjected to an external magnetic field applied along the tube axis \cite{portnoi2008terahertz,portnoi2009magnetically,hartmann2015terahertz} and therefore the results obtained herein are relevant to a broad range of quasi-one-dimensional Dirac systems.


\section{ACKNOWLEDGMENTS}
We are grateful to Charles Downing for the critical reading of the manuscript. This work was supported by the EU H2020 RISE project CoExAN (Grant No. H2020-644076), EU FP7 ITN NOTEDEV (Grant No. FP7-607521), FP7 IRSES projects CANTOR (Grant No. FP7-612285), QOCaN (Grant No. FP7-316432), and InterNoM (Grant No. FP7-612624). R.R.H. acknowledges financial support from URCO (Project No. 09 F U 1TAY15-1TAY16) and Research Links Travel Grant by the British Council Newton Fund.


\bibliography{Refs}

\end{document}